\documentstyle{article}

\def\ben{\begin{equation}}
\def\een{\end{equation}}
\def\bea{\begin{eqnarray}}
\def\eea{\end{eqnarray}}
\input amssym.def
\input amssym.tex
\begin{document}

\hfuzz=100pt
\title{Branes as BIons}
\author{G. W. Gibbons
\\
D.A.M.T.P.,
\\ Cambridge University, 
\\ Silver Street,
\\ Cambridge CB3 9EW,
 \\ U.K.}
\maketitle

\begin{abstract}
A BIon may be defined as  a finite energy
 solution of a non-linear field theory with distributional sources.
By contrast a soliton is usually defined to have  no sources.
I show how  harmonic coordinates map the exteriors of
the topologically and causally non-trivial
spacetimes of extreme p-branes to BIonic solutions
of the Einstein equations in a topologically trivial
spacetime in which the combined gravitational and matter energy momentum
is located on distributional sources. 
As a consequence the 
tension of  BPS  p-branes is classically
unrenormalized. The result holds equally for spacetimes with singularities
and for those, like the M-5-brane, which are everywhere singularity free.
\end{abstract}

One of the most striking aspects of the many recent applications
of p-branes  to black holes in
M-theory is the extent to which they admit two almost complementary
aspects. On the one hand one may view a p-branes as a flat
sheet-like
object of zero thickness moving in flat spacetime, described by 
a Dirac-Born-Infeld action, and 
on the other hand they may be regarded as curved spacetimes with
non-trivial topology and causal structure which solve the Einstein
equations \cite{GT,GHT}. This second aspect
 has become especially  prominent recently 
in the many papers in which the $AdS_{p+2} \times S^{d_T-1}$
geometry near the throat has played a vital role. In this paper
I wish to begin to address the question of why a description
of p-branes based on flat space can be so effective. This is of 
course part of a much wider puzzle; how is it that 
string theory and M-theory, based as they are on objects
moving in a fixed, and usually flat, background
give rise to theories like general relativity
in which no particular background is preferred?

If we view the problem from the point of
view of general relativity the answer is perhaps not so hard to see.
In  general relativity 
and related theories no particular coordinate
system is preferred and indeed it may be impossible to
find a single coordinate system
which covers the entire spacetime manifold ${\cal M}^n$.
However that does not prevent us fixing upon
a {\sl particular} set of coordinates $x^\alpha$ say and restricting
our coordinate transformations to Poincar\'e
transformations
of the $x^\alpha$. In 
other words we can always, locally at least,
 introduce an arbitrary flat spacetime
with inertial coordinates $x^\alpha$ and metric $\eta _{\mu \nu}$
and view gravity as the manifestation of a 
a non-linear spin two field in flat space.

In fact precisely this procedure is often
followed when one 
discusses the definition of energy and its conservation in 
general relativity \cite{P1,P2}. One assumes additionally
that the coordinates $x^\alpha$
are asymptotically Minkowskian in the sense that at large spatial distances
the curved spacetime metric $g_{\mu \nu}$ tends to the flat 
metric $\eta_{\mu \nu}$. Because one has a variational principle
one may construct the conserved canonical Einstein energy momentum
pseudo tensor $_E\frak{t}^\mu \thinspace _\nu$ such that the
conservation equation takes the form:
\ben
\partial _\mu \Bigl ( \sqrt{-g} T^\mu \thinspace _\nu + _E \frak{t}^\mu \thinspace _\nu
\Bigr )=0 \label{con}.
\een

The difficulty
is of course that the pseudo-tensor $_E\frak{t}^\mu \thinspace _\nu$ 
depends in an essential way on the chosen coordinates $x^\alpha$.
To some extent this does not matter
if one wishes to calculate for example, in an $E(p)$-invariant $(p+1+d_T)$-dimensional spacetime spacetime 
a quantities like the tension
\ben
T= - \int d^{d_T} y ( \sqrt{-g} T^0 \thinspace _0+ _E\frak{t}^0 \thinspace _0
\bigr )  
\een
because, as long as the coordinates
cover the whole spacetime ${\cal M}^n$,
this is independent of the choice of coordinates. 
However if the spacetime is not topologically trivial ${\cal M}^n \not \equiv {\Bbb R} ^n$ or if one seeks some more localized idea of energy one
has problems. 

One way out of this {\it impasse} is to fix the
troublesome gauge freedom once and for
all and to decree that although
all coordinates are equal some are more equal than others.
In other words, that, {\sl at least for some problems} ,
a particular choice of gauge is preferred.  This does not mean
giving up the equivalence principle  or the principle of
general covariance, any more than using
unitary gauge in Yang-Mills theory means
giving up Yang-Mills gauge invariance, it simply means
that in order to exploit  to the full our flat space
concepts we are going to pick the most convenient coordinate system
for that purpose.

This leads us to  the question: what is the most
convenient coordinate choice for studying p-branes? 
The suggestion of this paper is that the answer is {\sl harmonic
coordinates} for which:

\ben
\bigl (\sqrt{-g} g^{\mu \nu} \bigr ) _{,\mu}=0.
\een

Moreover I suggest that the most convenient choice
of variables to describe the gravitational field is  
\ben
\frak{g}^{\mu \nu} = \sqrt {-g} g ^{\mu \nu},
\een
in terms of which the harmonic gauge condition becomes
\ben
\frak{g}^{\mu \nu}\thinspace_{,\mu}=0. 
\een

Before discussing p-branes I will review some of the
(largely well known) properties of harmonic coordinates
and the variables $\frak{g}^{\mu \nu}$.
Firstly the name harmonic means just that: the $n$-functions
$x^\alpha$ are solutions of the curved space wave equation
\ben
\nabla ^2 x^\alpha { 1\over \sqrt{-g} } \partial _\mu \bigl 
( \frak{g}^{\mu \nu} x^\alpha _{, \nu} \bigr )=0.
\een
Another way of saying the same thing is that the identity map
is a harmonic map from $\{{\cal M}^n, g_{\mu \nu} \}$
to $\{ {\Bbb R}^n, \eta_{\mu \nu} \}$. In linear theory
the harmonic condition coincides with the De-Donder gauge
frequently used in perturbation theory.
This is because if $g_{\mu \nu} = \eta_{\mu \nu} + h _{\mu \nu}$ then
to lowest order
\ben
\frak{g}^{\mu \nu} \approx \eta ^\mu - h^{\mu \nu} + { 1\over 2} 
\eta ^{\mu \nu} h^\alpha _\alpha.
\een

Perhaps one of the most useful properties of the
gothic variables $\frak{g}^{\mu \nu}$ is that they behave nicely
under dimensional reduction. It is well known that if
a metric $g$ is the metric on a product
manifold :
\ben
g= g_1\oplus g_2
\een
one must Weyl rescale the metrics $g_1$ and $g_2$
to put them in Einstein conformal gauge. This is because
the actions don't add, in other words even though
\ben
R_{\mu \nu} (g)=R_{\mu \nu} (g_1)\oplus R_{\mu \nu} (g_2)
\een
it is not true that

\ben
\sqrt{-g} g^{\mu \nu}  R_{\mu \nu} (g) \ne  \sqrt{-g_1} g_1^{\mu \nu} R_{\mu \nu} (g_1)+ \sqrt{-g_2} g_2^{\mu \nu} R_{\mu \nu} (g_2) 
\een

However for the gothic variables if the metric is a product and
if in addition
\ben
\frak{g}^{\mu \nu}=\frak{g}_1^{\mu \nu}\oplus \frak{g}_2^{\mu \nu} \label{sum}
\een
then necessarily
 \ben
\frak{g}^{\mu \nu} R_{\mu \nu} (g) \ne \frak{g}^{\mu \nu} R_{\mu \nu} (g_1)+\frak{g}^{\mu \nu} R_{\mu \nu} (g_2). 
\een
 
Of course
\ben
\sqrt{-g_1} g_1^{\mu \nu} \ne \frak{g_1}^{\mu \nu}.
\een

In string theory products of metrics correspond to tensor
products of 
conformal field theories so the moral seems to be that
the use of the gothic variables better
respects that tensor product structure. As a practical matter it is certainly
easier to use (\ref{sum}) rather than to remember the formulae
for the Weyl rescaling.

It is of course standard that using the gothic variables 
$\frak{g}^{\mu \nu}$
simplifies the Lagrangian formulation.  If 
$\Gamma_\alpha \thinspace ^\beta \thinspace _\gamma = \Gamma_\gamma \thinspace ^\beta \thinspace _\alpha $ are the Christoffel symbols of the metric $g_{\mu \nu}$ then
one has the identity
\ben
\sqrt{-g} g^{\mu \nu} R_{\mu \nu} = \frak{g}^{\mu \nu} \bigl 
(\Gamma_\mu \thinspace ^\beta \thinspace _\alpha \Gamma_\nu \thinspace ^\alpha \thinspace _\beta-\Gamma_\mu \thinspace ^\alpha \thinspace _\nu
\Gamma_\alpha \thinspace ^\beta \thinspace _\beta  \bigl ) - \partial _\alpha \frak{ W } ^\alpha 
\een
where 
\ben
\frak{ W } ^\mu = \frak{g}^{\alpha \sigma} \Gamma _\sigma \thinspace ^\beta \thinspace _\beta - \frak{g} ^{\mu \nu}  \Gamma _\mu \thinspace ^\alpha \thinspace _\nu 
\een
Now if we define
\ben
\frak{L}= { 1\over 16 \pi} \frak{g}^{\mu \nu} \bigl 
(\Gamma_\mu \thinspace ^\beta \thinspace _\alpha \Gamma_\nu \thinspace ^\alpha \thinspace _\beta-\Gamma_\mu \thinspace ^\alpha \thinspace _\nu
\Gamma_\alpha \thinspace ^\beta \thinspace _\beta  \bigl ),
\een
then we find that that $\frak{L}$ contains no second derivatives
of and is  a homogeneous function of degree
$-1$ in $ \frak{g}^{\mu \nu}$ and a homogeneous function of degree $2$
in $ \frak{g}^{\mu \nu}\thinspace _{,\lambda}$.
Moreover
\ben
\frak{W}^\alpha = { 1 \over 16 \pi} \frak{g}^{\mu \nu} { \partial \frak{L} \over \partial 
\frak{g}^{\mu \nu}\thinspace _{,\alpha}} \label{noether}.
\een
It follows from (\ref{noether}) that the current density $\frak{W}^\alpha$
has the interpretation of the Noether current associated to dilations or homotheties.
Rescaling the coordinates $x^\alpha$ is equivalent to rescaling the 
gothic variable $\frak{g}^{\mu \nu}$. The associated Noether charge
is of course closely related to the surface term in the gravitational action.  
We shall not pursue that avenue here but instead remark that 
the Einstein pseudo-tensor $\frak{t} ^\mu \thinspace _\nu$ which appears
in (\ref{con}) is just
the canonical energy-momentum tensor associated to $\frak{L}$, i.e.
\ben
_E\frak{t}^\mu \thinspace _\nu= \delta ^\mu_\nu \frak{L} - \frak{g}^{\alpha \beta}\thinspace _{,\nu} { \partial \frak{L} \over \partial 
\frak{g}^{\alpha \beta}\thinspace _{,\mu}}.
\een

Its conservation (\ref{con}) may be ascribed to
the invariance of the action associated to the
Lagrangian $\frak{L}$ under translations
along the directions of the arbitrary coordinate chart $\{x^\alpha \}$.
Of course to obtain something like the conventional idea of energy
one chooses the chart $\{x^\alpha\}$ to be asymptotically Minkowskian.
If the spacetime manifold $M^n$ is not topologically trivial then
one will not be able to define the translations globally all over $M^n$.
This is of course the problem one faces if black holes or branes are present.
In the case of extreme branes however, and for the particular choice of
harmonic coordinates, we shall see that this problem is 
somewhat alleviated since the exterior region of the branes
is mapped to ${\Bbb R}^{p+1}  \times ( {\Bbb R} ^{d_T} \setminus \{ {\bf y}_i\} )
$ where $\{ {\bf y}_i \}$ are the positions of the branes.

The Einstein pseudo-tensor $_E\frak{t}^\mu _\nu$ has a number of 
counter-intuitive properties. For example Schr\"odinger \cite{S} has pointed out
that if one calculates it for the Schwarzschild solution using the 
coordinates $ x^\alpha = (t, r \sin \theta \cos \phi , r \sin \theta \sin \phi, r \cos \theta )$, where $r$ is the usual Schwarzshild radial coordinate defined so that
the area of a sphere is $4 \pi r^2$,
one finds that it vanishes, at least away from the horizon at $r=2M$
and the singularity $r=0$ , where it is not defined. The coordinates chosen by Schr\"odinger have the property that $\sqrt{-g}g=1$ but they are not 
in fact harmonic. One obtains harmonic coordinates by taking
$ x^\alpha = (t, (r-M) \sin \theta \cos \phi , (r-M) \sin \theta \sin \phi, (r-M) \cos \theta )$. In fact this formulae gives harmonic coordinates for
the entire Reissner-Nordstrom family. For later use we note that
isotropic coordinates $ x^\alpha = (t, \rho \sin \theta \cos \phi , \rho \sin \theta \sin \phi, \rho \cos \theta )$ where
\ben
r= \rho + M + { M^2 - Z^2 \over 4 \rho}
\een and $Z^2= q^2 + p^2$ are harmonic if and only if 
\ben
M=\pm |Z|.
\een
Of course we usually take the plus sign to get a solution
with a horizon rather than one with a naked singularity
but as with the condition for
the existence of Killing spinors either sign is actually allowed.

Another disadvantage of the Einstein pseudo-tensor is that
it does not give rise to  simple expressions for the total angular momentum.
This is because if one raises an index with $g^{\mu \nu}$ it is not 
necessarily symmetric. Thus if one introduces a flat metric $\eta _{\mu \nu}$
and uses it to define a set of lorentz transformations
with respect to the chart $\{ x^\alpha\}$ one
will not obtain directly form $_Et^\mu _\nu$ a set of conserved currents.
It is clear however by Noether's 
theorem that such currents can always
be constructed. Indeed
associated with any one parameter family of diffeomorphisms 
there exists such a current. In fact these currents are not unique.
One possible choice is that of Landau and Lifshitz. They introduce
a symmetric \lq complex" $_{LL}\frak{t}^{\mu \nu}= _{LL}\frak{t}^{\nu \mu}$ in terms of the  the symmetric quantity
\bea
_{LL}\Theta ^{\mu \nu} &=& \sqrt{-g}  \bigl ( \frak{T}^{\mu \nu} + \frak{t} ^{\mu \nu} \bigr ) \cr \nonumber
&=& \partial _ \beta \partial _\alpha \Bigr ( { 1 \over 16 \pi } 
\bigl ( \frak{g}^{\mu \nu}\frak{g}^{\alpha \beta}-\frak{g}^{\mu \alpha}\frak{g}^{\beta \nu} \bigr ) \Bigr ) \label{LL}
\eea
It follows that
\ben
\partial _\mu { _{LL} \Theta ^{\mu \nu} } =0 \label{newcon}.
\een
 
One may therefore, with the same caveats as before, regard $_{LL} \Theta ^{\mu \nu}$
as giving the distribution of  energy and momentum 
for the combined gravitational and matter fields. As with the Einstein-pseudo-tensor,
so with the Landua-Lifshitz complex, it is reallly only integral 
quantities for asymptotically flat spacetimes 
which are invariant under change of the coordinate chart $\{x^\alpha \}$. 

For static $E(p,1)$-invariant asymptotically flat brane configurations
multiplication of (\ref{newcon}) by $x^\alpha$ and integration by parts
leads to some non-trivial Virial relations for the 
integrals of the non-zero components of $ _{LL} \Theta ^{\mu \nu}$
over the transverse dimensions. Thus
\ben
\int d^{d_T} y _{LL} \Theta^{ij} =0, 
\een
and
\ben
\int d^{d_T} y _{LL} \Theta^{ab} =\eta ^{ab} T,
\een
where $i,j=i,\dots, d_T $ and $a,b= 0, \dots, p$ and $T$ is the tension,
i.e. the energy per unit p-volume.

In harmonic coordinates the formula for $ _{LL} \Theta ^{\mu \nu}$
simplifies
\ben
 _{LL} \Theta ^{\mu \nu}= 
{ 1 \over 16 \pi} \Bigl ( \frak{g}^{\alpha \beta}
\partial _\alpha \partial _\beta \frak{g}^{\mu \nu}- 
\partial_\alpha  \frak{g}^{\beta \nu} \partial_\beta  \frak{g}^{\mu \alpha} \bigl ) \label{formula}.
\een

The harmonic function rule for orthogonally
\lq intersecting" or \lq overlapping "
branes depending on a set of harmonic functions
$H_i({\bf y})$ on ${\Bbb E} ^{d_T}$ states that 
\begin{itemize}
\item
 the metric is diagonal,
\item
that while the transverse space is not flat
nevertheless
\ben
\frak{g}^{ij}= \delta ^{ij}.
\een
\item
The time direction is  common to all the branes and 
\ben
\frak{g}^{00}=- \prod H_i
\een
\item In a direction in one or more branes 
\ben
\frak{g}^{aa} = \prod _{a} H_i,
\een
where  $\prod _{a} { }$ denotes a product over all
the harmonic functions associated with branes
sharing the direction $a$.
\end{itemize}
A simple calculation reveals that the coordinates
$x^\alpha = (x^a, y^i)$ are harmonic. Moreover in the case
of a single type of brane, with just one harmonic function,
corresponding to branes located at positions  ${\bf y}= {\bf y}_i \in {\Bbb R}^ {d_T}$,
the transverse stresses vanish point-wise
\ben
_{LL} \Theta^{ij}=0.\label{stress},
\een
while the energy-momentum is strictly localized on the branes
\ben
_{LL} \Theta^{ab}=- \eta ^{ab} \sum _i T_i \delta ({\bf y}_i) \label{energy}.
\een
For more than one type of brane
the property (\ref{stress}) that the stresses
vanish remains true but the distribution of energy-momentum
is more complicated than (\ref{energy}). This is presumably not unrelated to the fact while the single branes cary no entropy, sytems of intersecting
branes can.

These results apply in particular to the 
both the \lq elementary" M-2-brane and the \lq solitonic"
M-5-brane. This is quite surprising. Both
have a non-trivial topological and causal structure \cite{GT, GHT,DGT}.
The former has Reissner-Nordstrom like geometry with a singularity
inside an event horizon, the latter is everywhere non singular \cite{GHT}.
Harmonic coordinates map the exteriors
of both to ${\Bbb R}^{11}$ with distributional
sources, a fact first noted for the M-2-brane
by Duff and Stelle \cite{DS}. As they observed, double-dimensional
reduction yields the fundamental string in 10 dimensions
\cite{DH,DGHR} which is truly
singular but has a distributional source. As noted in \cite{DH}
and \cite{DGHR} the string tension is not renormalized.
This is of course consistent with the present analysis and gibes
with    
some recent work on cosmic strings \cite{BC,AD}.
These papers consider time-dependent cosmic strings in
first order perturbation theory. In the present paper I have considered
static branes in the exact theory. They find that self-interactions
do not result in a classical renormalization of the string tension
and that in four spacetime dimensions gravitational 
self-interactions vanish. 
At the linear level this cancellation
is a direct consequence of the BPS
condition and is closely related to the antigravitating
properties of the solutions, since the
sum of the relevant propagators vanishes
as a consequence the antigravity condition cf \cite{H}.
Similar observations may be found in the 
old literature on self-energies \cite{SS}.
The present paper establishes
the classical non-renormalization property for general $p$
in the fully non-linear theory.

In addition to branes, one frequently considers waves
in ${\Bbb R} ^2 \times {\Bbb R}^{d_T} $ with coordinates
$x^+, x^-, {\bf y}$. Suppose that the wave moves along the 
$x^-$ direction then
\ben
\frak{g}^{ij}= \delta ^{ij}
\een
\ben
\frak{g}^{+-}= 1
\een
\ben
\frak{g}^{++}= -F(x^-, {\bf y}). 
\een
Evidently the coordinates $\{x^+,x^-, {\bf y}\}$ are harmonic
irrespective of the precise form of the function $F(x^-, {\bf y})$
as is the vanishing stress condition (\ref{stress}).
In fact the only non-vanishing component of $ _{LL} \Theta ^{\mu \nu}$
is
\ben _{LL} \Theta ^{++}={ 1\over 16 \pi} \nabla ^2 F\label{wave}.
\een
For a non-singular pure gravitational wave, 
the function $F(x^-, {\bf y})$ is everywhere non-singular and the right hand side of (\ref{wave}) vanishes.
This does not invalidate the idea of using non-tensorial
measures of energy and momentum because a plane wave spacetime is not 
asymptotically flat but it does illustrate the potential hazards.
For the singular waves that give rise on dimensional reduction
to $0$-branes, or indeed higher dimensional branes,
one choses $F$ to be independent of $x^-$ and finds
that there is
is a distributional source just as in (\ref{energy}).

One may how the results of this paper are
altered  if one chooses
a different expression for the local density of energy and momentum.
The present observations were in fact stimulated by reading a paper
of Papapetrou \cite{P1} who also used harmonic coordinates 
and who explicitly introduced a flat metric $\eta ^{\alpha \beta}$.
His  expression, call it $_P\Theta ^{\mu \nu}$,
for the local density of energy and momentum reduces
in harmonic coordinates to 
\ben
_P \Theta ^{\mu \nu}= \Box \frak{g}^{\mu \nu},
\een
where $ \Box = \eta ^{\alpha \beta }\partial _\alpha \partial _\beta $
is the flat space D'Alembertian operator. In his later textbook
\cite{P2} he abandons his earlier
approach and instead treats the Landau-Lifshitz formalism.
Applied to M-branes one gets the same results with either formalism.
 
In a recent paper on Born-Infeld theory \cite{G1}
the concept of a BIon was introduced.
This is a finite energy solution of a non-linear theory with a 
distributional source. In the case of the standard Born-Infeld
theory, there is a source of electric charge and the analogue
of (\ref{energy}) is  
\ben
J^0 = { 1\over 4 \pi} {\rm div} {\bf D}=\sum _i  e_i \delta({\bf y}_i) \label{Gauss},
\een
where the zeroth componet of the current
 $J^0$ is the analogue of $_{LL}\Theta ^{00}$.
In a sense the results of the present paper may be paraphrased
by saying that harmonic coordinates map extreme p-branes into
a  special kind of  BIon associated to a non-linear spin two
field. In the case of the non-linear theory
of spin one, Gauss's theorem 
says that despite the polarization of the
vacuum,  the  electric charge of a BIon is
not classically renormalized. In the case of the gravity the tension is not classically renormalized.

\end{document}